\documentclass[prl,twocolumn,superscriptaddress]{revtex4}
\usepackage{amssymb,amsmath,amsthm,graphicx}
\usepackage{latexsym}
\usepackage{bm}
\usepackage[]{hyperref} % to have online links

\def\be{\begin{equation}}
\def\ee{\end{equation}}
\def\beq{\begin{eqnarray}}
\def\eeq{\end{eqnarray}}

\pdfoutput=1

\begin{document}

\def\lsim{\mathrel{\rlap{\lower4pt\hbox{\hskip1pt$\sim$}}
    \raise1pt\hbox{$<$}}}
\def\gsim{\mathrel{\rlap{\lower4pt\hbox{\hskip1pt$\sim$}}
    \raise1pt\hbox{$>$}}}
%\def\sqr#1#2{{\vcenter{\vbox{\hrule height.#2pt
%         \hbox{\vrule width.#2pt height#1pt \kern#1pt
%         \vrule width.#2pt}
%         \hrule height.#2pt}}}}
%\def\square{\mathchoice\sqr66\sqr66\sqr{2.1}3\sqr{1.5}3}
%%%%%%%%%%%%%%%%%%%%%%%%%%%%%%
\def\be{\begin{equation}}
\def\ee{\end{equation}}
\def\bea{\begin{eqnarray}}
\def\eea{\end{eqnarray}}
%%%%%%%%%%%%%%%%%%%%%%%%%%%%%%
\newcommand{\der}[2]{\frac{\partial{#1}}{\partial{#2}}}
\newcommand{\dder}[2]{\partial{}^2 #1 \over {\partial{#2}}^2}
\newcommand{\dderf}[3]{\partial{}^2 #1 \over {\partial{#2} \partial{#3}}}
\newcommand{\eq}[1]{Eq.~(\ref{eq:#1})}
\newcommand{\dd}{\mathrm{d}}

\title{The instability of orientifolds and the end of the (string) landscape as we know it}

\author{Keith Copsey} 
\affiliation{ Perimeter Institute for Theoretical Physics, Waterloo, Ontario N2L 2Y5, Canada\\
\&  Department of Physics \& Astronomy, University of Waterloo, Waterloo, Ontario N2L 3G1, Canada}
\email{kcopsey@perimeterinstitute.ca}

\begin{abstract}

Negative mass brane-like objects known as orientifolds are crucial in the construction of all presently understood realistic string compactifications and I review in general the link between negative energy objects and the existence of a landscape of metastable vacua.  While in string perturbation theory these solutions are non-dynamical, and hence avoid the usual instabilities of negative tension objects, beyond perturbation theory it is well known that one may lift orientifolds in M-theory to solutions well described by classical general relativity and which consequently may be distorted in a variety of ways.    I review the failure of the usual positive energy theorems, as well as several supersymmetric stability arguments, in this context and demonstrate that one may smoothly distort the lifted orientifold, preserving the boundary conditions, into solutions of lower energy.  I present significant (numerical) evidence that solutions with arbitrarily negative energy may be found among these distorted orientifolds.  Consequently, I suggest that this instability is a fatal one and hence there is little reason to believe in the existence of a string theory landscape.   I briefly comment on the implications for Type I string theory and string dualities.
\end{abstract}

\maketitle

%%%%%%%%%%%%%%%%%%%%%%%%%%%%%%%%%%%%%%%%%%%%%%%%%%%%%%%%%%%%%%%%%%%%%%%%%%%
%%%%%%%%%%%%%%%%%%%%%%%%%%%%%%%%%%%%%%%%%%%%%%%%%%%%%%%%%%%%%%%%%%%%%%%%%%%

It has become standard lore that string theory predicts a vast array of vacua, numbering apparently at least of order $10^{500}$, known as the landscape constructed by arranging fluxes, branes, and other stringy contributions in various configurations in compact extra dimensions.   Physics at energies small compared to the Planck scale may be expected to be wildly different in these different vacua and without further input it appears one can do no better than make probabilistic predictions, each of which suffers from a measure ambiguity regarding which one may count on one hand the number of individuals who agree on a particular approach.   On the other hand, working out the detailed physical predictions of any particular vacua turns out to be a highly complicated effort that often may require a year's worth of work and finding any vacua which allows the Standard Model of particle physics in detail and does not contradict basic cosmological observations remains an unresolved challenge despite the dedicated efforts of capable individuals.  In view of these difficulties it might strike one that perhaps the foundations of this framework deserve reconsideration.  Indeed, had one approached one's friendly neighbourhood relativist in the eighties and suggested one could construct not just one or two but a vast number of static (or nearly so) solutions by arranging Planck-scale fluxes and other matter contributions in various configurations in compact extra dimensions the response almost certainly would have been that such a scenario appeared ludicrous-unless one had some source of large negative energy density, in which case the theory was presumably unstable, the fluxes would result in a Planck scale effective cosmological constant, at least in the large if not the compact dimensions, resulting in a highly dynamical solution that, presuming it did not immediately collapse into a singularity, was in obvious contradiction with cosmological observations.    

In the case of string theory, the above mentioned source of negative energy density is provided by orientifolds, brane-like configurations which have negative tension and at least in string perturbation theory (on a fixed background) do not couple to the graviton and hence are non-dynamical.  It may be worth emphasizing that while there are string compactifications that do not involve orientifolds, ranging from the trivial toroidal compactification to the much more realistic hetereotic configurations, to the best of my knowledge to date no compactification has succeeded in stabilizing all moduli and breaking supersymmetry in a controlled fashion without introducing the need for fluxes or other positive energy contributions and hence a need for negative energy sources as discussed above.  The orientifold combines an orbifold, in particular a $Z_2$ reflection symmetry, with a worldsheet parity reversal which reverses the orientation of strings.  A negative tension brane (i.e. one whose mass decreases as it is stretched) would appear disastrous but a non-dynamical source of negative energy, like a negative cosmological constant, need not imply any instability. On the other hand, beyond perturbation theory on a fixed spacetime background, it is difficult to reconcile concepts like a dynamical spacetime and causality with a fundamentally non-dynamical object, especially in a theory like string theory that reproduces general relativity at low energies.  In particular, it is known that the an O6-plane, namely an orientifold with six flat dimensions, may be lifted in M-theory to a purely gravitational regular Ricci-flat solution consisting of a product manifold of seven dimensional Minkowski space and a Ricci-flat solution known as the Atiyah-Hitchin metric with negative ADM mass (\cite{MarolfRossOrient}-\cite{Dorey}) (for a particularly concise and cogent review see \cite{MarolfRossOrient}).   For large string coupling, this lifted solution is well described as a purely gravitational solution in general relativity in eleven dimensions so one may then freely consider various distortions, generically dynamical, of the orientifold and see whether one might find other solutions smoothly connected to the orientifold  with the same boundary conditions (and topology) with a more negative ADM mass, as the above concerns suggest.  Note since T-duality connects the Op-planes with various p, just as it does D-branes, one looses nothing by focusing on the 06-plane.

The Atiyah-Hitchin metric may be written in terms of the left-invariant one-forms on $S_3$ as

\be \label{met1}
ds_{AH}^2 = f^2(r) dr^2 + a^2 (r) \sigma_1^2 + b^2(r) \sigma_2^2+c^2(r) \sigma_3^2
\ee
where
\begin{eqnarray}
\sigma_1 &=& - \sin \psi d\theta + \cos \psi \sin \theta d\phi,
\nonumber \\ 
\sigma_2 &=& \cos \psi d\theta + \sin \psi \sin \theta d\phi,
\nonumber \\
\sigma_3 &=& d \psi + \cos \theta d\phi,
\end{eqnarray}
with $0 \leq \theta \leq \pi$, $0 \leq \phi \leq 2 \pi$, and $0 \leq \psi \leq 2 \pi$ and the functions of $r$ are specified via
\be \label{eqncyclic}
\frac{2 b(r) c(r) a'(r)}{f(r)} = \Big(b(r)-c(r)\Big)^2 - a^2(r)
\ee
and the two equations related to (\ref{eqncyclic}) to cyclically permuting $a(r)$, $b(r)$, and $c(r)$.   The precise Atiyah-Hitchin solution, as well as the lift of the O6-plane, also has two discrete symmetries imposed upon it; I: ($\theta \rightarrow \pi - \theta, \,  \phi \rightarrow \pi + \phi, \, \psi \rightarrow -\psi$) and II: ($\theta \rightarrow \theta, \, \phi \rightarrow \phi, \,  \psi \rightarrow \pi + \psi$) or in terms of the above one forms I: ($\sigma_1\rightarrow \sigma_1, \, \, \sigma_2\rightarrow -\sigma_2, \, \, \sigma_3 \rightarrow -\sigma_3$) and  II: ($\sigma_1\rightarrow -\sigma_1, \, \, \sigma_2\rightarrow -\sigma_2, \, \, \sigma_3 \rightarrow \sigma_3$) .  The asymptotics resemble Kaluza-Klein, where in the most natural gauge at large $r$, $f(r) \rightarrow 1$, $a(r) \rightarrow r$, $b(r) \rightarrow r$ and $c(r) \rightarrow -R_{11}$ where $R_{11}$ is the size of the M-theory circle.    In particular, if one chooses the gauge $f(r) = -b(r)/r$, up to corrections exponentially suppressed at large $r$, $a(r) = b(r) = r \sqrt{1-R_{11}/r}$ and $c(r) = -R_{11} /\sqrt{1-R_{11}/r}$ which the reader may recognize as the metric for negative mass Taub-Nut.    Since $\sigma_1^2 + \sigma_2^2 = d\theta^2 + \sin^2 \theta d \phi^2$ one ends up with a space that is asymptotically  $R^3 \times S_1/\mathcal{Z}_2$, where the $S_1$ comes from the $\sigma_3$ direction and the $\mathcal{Z}_2$ is the identification II.  If one writes  ($x = r \sin \theta \cos \phi, y = r \sin \theta \sin \phi, z = r \cos \theta$), then the identification I. becomes $(x,y, z) \rightarrow (-x,-y,-z)$, in other words the orbifolding part of the orientifold. 

As one goes to $r$ comparable to $R_{11}$ the asymptotically exponentially small corrections smooth out the singularity of negative mass Taub-Nut and one finds instead a bubble of nothing or bolt, depending on one's favorite terminology, as $a(r)$ goes to zero at a finite $r$.  In particular, again in the gauge $f(r) = -b(r)/r$ near $r_0 =  \pi R_{11}/2$, $a(r) = 2 (r -r_0) + \mathcal{O}((r-r_0)^2), b(r) = r_0 + (r-r_0)/2+ \mathcal{O}((r-r_0)^2),  c(r) = -r_0 + (r-r_0)/2+ \mathcal{O}((r-r_0)^2)$.  This smoothly pinches off the $\sigma_1$ direction and results in a metric which is regular and geodesically complete.  To see this explicitly, it is useful to write the one-forms in an alternate coordinate system from the above 
\begin{eqnarray}
\sigma_1 &=& d \bar{\psi} + \cos \bar{\theta} d\bar{\phi}
\nonumber\\
\sigma_2 &=& - \sin \bar{\psi} d\bar{\theta} + \cos \bar{\psi} \sin \bar{\theta} d\bar{\phi}
\nonumber \\ 
\sigma_3 &=& \cos \bar{\psi} d\bar{\theta} + \sin \bar{\psi} \sin \bar{\theta} d\bar{\phi}
\end{eqnarray}
and then the Atiyah-Hitchin metric near $r= r_0$ may be written as
\be \label{nearmet}
ds^2 \approx dr^2 +4 (r-r_0)^2 (d\bar{\psi} + \cos \bar{\theta} d \bar{\phi})^2 + r_0^2(d\bar{\theta}^2 + \sin^2\bar{\theta} d\bar{\phi}^2)
\ee
Ignoring identifications for a moment, the first terms in the metric become $R_2$ while the second two become $S_2$.  The identification I. means the period of $\bar{\psi}$ is $\pi$, ensuring we do not have a conical singularity as $\bar{\psi}$ pinches off, while the identification II., while not necessary for a regular solution, converts the latter part of the metric from an $S_2$ into $RP_2$.  

Note in eleven dimensions the solution is regular and purely gravitational, whereas if one tried to analyze the system in ten dimensions one would encounter what looks like a naked singularity (disturbingly similar to the negative mass Schwarzschild black hole) and Kaluza-Klein gauge fields (including massive KK modes which act like negative energy matter violating the usual energy conditions \cite{MarolfRossOrient}).   However, due to the above identifications, the Atiyah-Hitchin metric is only locally asymptotically flat and lacks a global covariantly constant spinor  \cite{GibbonsSUSYDomainWall}  of the type needed to prove a positive energy theorem along the lines pioneered by Witten \cite{Wittenposenergy}.   Indeed the Atiyah-Hitchin metric is just one example of a variety of examples of locally asymptotically flat or locally asymptotically Euclidean solutions where one can not prove a positive energy theorem and where one can indeed find negative energy states \cite{ALFexamples}.  From the perspective of the SUSY algebra, such a conclusion is essentially inevitable; if one could define the supersymmetry generators globally the Hamiltonian would be given by anticommuting a SUSY generator with its conjugate and consequently, as usual, be positive definite while the lift of the negative tension orientifold must have negative mass (i.e. a smaller mass than orbifolded flat space).  It may also be worth emphasizing that the existence of an asymptotic supersymmetry is not sufficient to prove a lower bound to the energy--a negative mass black hole being probably the simplest example-- and one needs some proof that extends into the interior of the spacetime, as the Witten argument does.   More generally, in the presence of gravitational backreaction even if one's initial state is (globally) supersymmetric it need not be energetically stable--a generic perturbation to the state will make it non-supersymmetric and that perturbed state need not be stable in any sense, as made famous by Coleman and de Luccia \cite{CdL} in the discussion of flat space into anti de Sitter.  Consequently we seem to have no good reason to believe the lifted orientifold is energetically stable.  Recalling also the fact that asymptotically Kaluza-Klein bubble of nothing solutions have been known to result in solutions with arbitrarily negative energy \cite{KKbubbles}, at least in the absence of a stabilizing effect like the confining ``box''  of an asymptotically AdS spacetime \cite{AdSSoliton,CopseyHorowitzsol}, it appears there is every reason to suspect one might be able to find solutions with masses more negative than the Atiyah-Hitchin solution.

In the search for lower mass solutions, limiting ourselves to static solutions would be only a rather limited exploration of this question, especially since one does not expect members of a family of increasingly negative energy solutions to be dynamical--given any member of such a family one would expect it to evolve towards another member of the family with lower energy, dynamically acquiring along the way enough gravitational momentum to conserve energy.   Since one expects neither gravitational momentum nor distorting the flat directions will make the energy more negative, let us consider initial data with no initial gravitational momentum and a spatial metric which leaves the flat directions alone but distorts the Atiyah-Hitchin metric ($\times R^{(6)}$) 
\be \label{met2}
ds^2 = dr^2/W(r) + a^2 (r) \sigma_1^2 + b^2(r) \sigma_2^2+c^2(r) \sigma_3^2
\ee
where now we choose the above functions of $r$ as we wish, maintaining the regularity and boundary conditions above, subject to the constraints, that is the components of Einstein's equations involving no second time derivatives which must be imposed on one's solution at any given time.  The remainder of the Einstein equations then act as evolution equations.  Note also the given ansatz (\ref{met2}) respects the projections I and II discussed above for any choices of $(a(r), b(r), c(r))$ whereas any term mixing different 1-forms (i.e. $f(r) \sigma_1 \sigma_2$) would be forbidden by these symmetries.  At the risk of belaboring the obvious, one has in mind of course not a truly infinite $R^{(6)}$ but as is customary for D-branes and orientifolds wrapping these directions on some compactification (e.g. a torus) of finite volume.  For the case of present interest, the Hamiltonian scalar constraint is the only nontrivial condition (see, e.g. \cite{ReggeTeitelboim, SudarskyWald,CopseyMann}) namely that the curvature constructed from the spatial metric vanishes.  Since the flat directions contribute nothing to the curvature, this is just the statement that the curvature of the metric (\ref{met2}) vanishes, or
\begin{eqnarray} \label{Const}
0 &= &R^{(10)} = R^{(4)} =  -\Big(\frac{a'(r)}{a(r)}+\frac{b'(r)}{b(r)}+\frac{c'(r)}{c(r)}\Big)W'(r)\nonumber \\
&-& \Big(\frac{2 a'(r) b'(r)}{a(r) b(r)} + \frac{2 a'(r) c'(r)}{a(r) c(r)} +\frac{2 b'(r) c'(r)}{b(r) c(r)} +\frac{2 a''(r)}{a(r)} \nonumber \\
 &+& \frac{2 b''(r)}{b(r)}+ \frac{2 c''(r)}{c(r)} \Big) W(r) \nonumber \\
 &-& \frac{a^4(r) - 2 a^2(r) ( b^2(r) + c^2(r)) +(b^2(r) - c^2(r))^2}{2 a^2(r) b^2(r) c^2(r)}
\end{eqnarray}
One may then specify $a(r)$,$b(r)$, and $c(r)$ and (\ref{Const}) is simply a first order ordinary differential equation for $W(r)$ for which one may, as usual, write a solution in terms of double integrals.   As long as one chooses functions $(a(r),b(r),c(r)$) which are $C_1$ then W(r) will be continuous, as one may convince oneself just by staring at (\ref{Const}) for a moment or, if one prefers, by solving this ODE and writing the solution for $W(r)$ in a form involving only the functions and their first derivatives.   This double integral structure, however, makes it difficult to obtain non-trivial analytic results and I will largely confine my attention to numerical solutions of (\ref{Const}). 

For the given asymptotics, the ADM mass, or more properly the on-shell value of the Hamiltonian, derived from a Lagrangian $\mathcal{L} = \kappa \int \sqrt{-g} R$ is given \cite{ReggeTeitelboim, SudarskyWald,CopseyMann})for a spatial metric $h_{a b}$ which approach the asymptotic (spatial) metric $h^{(0)}_{a b}$ by
\be
E =  \kappa \int_{d\Sigma} dS^{a} \Big[ D^b (\delta h_{a b}) -  D_a \delta h   \Big]
\ee
where indices are raised and lowered with $h^{(0)}_{a b}$, $D_a$ the covariant derivative compatible with $h^{(0)}_{a b}$, $\delta h_{a b} = h_{a b} - h^{(0)}_{a b}$,  $\delta h = {h^{(0)}}^{a b} \delta h_{a b}$ and the integral is performed over the spatial infinity for one's manifold $d\Sigma$. Provided that asymptotically the functions in the metric (\ref{met2}) approach
\begin{eqnarray} \label{asymmet}
W(r) &=& 1 - \frac{\mu(r)}{r}, \, \, \, c(r) = -R_{11} \Big(1+ \frac{\delta c(r)}{r} \Big), \nonumber \\
a(r) &=& r \Big(1+ \frac{\delta a(r)}{r} \Big) \, \, \,  b(r) = r \Big(1+ \frac{\delta b(r)}{r} \Big)
\end{eqnarray}
where $\mu(r)$ and $(\delta a(r), \delta b(r), \delta c(r))$ have finite limits and  $(r \delta a'(r), r \delta b'(r), r \delta c'(r))$ vanish asymptotically then the Hamiltonian is well-defined and one has an energy
\be
E =  4 \pi^2 \kappa R_{11} V_{T} \Big( \mu(\infty) + \delta c(\infty) \Big)
\ee
where $V_{T}$ is the volume of the transverse $R^{(6)}$ directions.  In particular given the asymptotics above, the energy of the Atiyah-Hitchin solution is $E_{AH} = -2 \pi^2 \kappa R^2_{11}V_T$.

Let us consider a family of what one might call tri-axial distortions of the Atiyah-Hitchin metric by taking
\begin{eqnarray} \label{Confdef}
a(r) &=& \gamma_a(r) a_0(r), \, \, b(r) = \gamma_b(r) b_0(r), \, \, \,  c(r) = \gamma_c(r) c_0(r) \nonumber \\
W(r) &=& \frac{r^2}{b^2(r)} \Big(1 - \frac{\mu_0(r)}{b(r)} \Big)
\end{eqnarray}
and $(a_0(r), b_0(r), c_0(r))$ are the functions in the original Atiyah-Hitchin metric, i.e.
\be \label{eqncyclic2}
\frac{2 b_0(r) c_0(r) a_0'(r)}{f_0(r)} = \Big(b_0(r)-c_0(r)\Big)^2 - a_0^2(r)
\ee
plus cyclic permutations and, as before, we will choose the gauge $f_0(r) = -b_0(r)/r$ for later convenience.    If we took $\gamma_a(r_0) =  \gamma_b(r_0) =\gamma_c(r_0) =1$ then we would recover the Atiyah-Hitchin solution;  it is easy to check in this case the constraint (\ref{Const}) implies that a nonzero $\mu_0(r)$ would imply $\mu_0(r_0)$ is nonzero which would, as will be discussed shortly, in turn implies the existence of a conical singularity.   More generally, as long as we consider distortions of these $\gamma_i$ which are regular and nonzero everywhere the squashed solution will have the same topology as Atiyah-Hitchin.  Requiring that $W(r)$ approaches a constant at large $r$ implies via the constraint that asymptotically $\gamma_a(r) \rightarrow \gamma_b(r)$; otherwise one finds a de Sitter like form $W(r) \sim -r^2$.  Defining $\Lambda = \gamma_a(\infty) = \gamma_b(\infty)$ and $\Lambda_1 = \gamma_c(\infty)$, then at large $r$, $c^2(r) \rightarrow R_{11}^2 \Lambda_1^2$ and $a^2(r) \rightarrow b^2(r) \rightarrow  \Lambda^2 r^2$.   Defining the new coordinate $\bar{r} = \Lambda r$ and physical size of the M-theory circle $\bar{R}_{11} = \Lambda_1 R_{11}$, one obtains asymptotics as the same form as before (\ref{asymmet}).  Specifically if asymptotically
\begin{eqnarray}
 \gamma_a(r) &=& \Lambda \Big(1+ \frac{\delta \gamma_a(r)}{r}\Big),  \gamma_b(r) = \Lambda \Big(1+ \frac{\delta \gamma_b(r)}{r}\Big) \nonumber \\
  \gamma_c(r) &=& \Lambda_1 \Big(1+ \frac{\delta \gamma_c(r)}{r}\Big) \
\end{eqnarray}
where each of the $\delta \gamma_i$'s has a smooth limit (in particular requiring $r \delta \gamma_i' \rightarrow 0 $ asymptotically)  then one has an energy
\be \label{E2}
E = 4 \pi^2 \kappa \bar{R}_{11} V_T \Big(\mu_0(\infty) + \Lambda[2 \delta \gamma_b(\infty)+\delta \gamma_c(\infty) ] -\frac{\Lambda}{2\Lambda_1} \bar{R}_{11} \Big)
\ee
It may be worth  emphasizing that while (\ref{E2}) is written in terms of the more convenient r (i.e. $\mu_0(\infty) = \lim_{r \rightarrow \infty} \mu_0(r) $) the energy has been calculated by transforming to standard coordinates $(\bar{r}, \bar{R}_{11})$ to avoid large coordinate transformation issues.  However, despite their explicit appearance in (\ref{E2}) as a matter of practice there does not appear to be any particular advantage in looking for low energy solutions to considering solutions where $\Lambda$ or $\Lambda_1$ differs from unity; upon requiring interior regularity the constraint tends to enforce a $\mu_0(r)$ that at least compensates for any advantage one might gain by taking, for example, $\Lambda/\Lambda_1 \gg 1$.  One effect of taking $\Lambda \neq \Lambda_1$ is to change the first order corrections to $(a(r), b(r), c(r))$  (e.g. $\delta c(r)$ in \ref{asymmet}) and while the usual gravitational definitions of normalizability (i.e. finiteness of the Hamiltonian) would seem to allow such changes, it is not inconceivable that one might be able and want to enforce a stronger boundary condition.  To alleviate any such concerns, I will focus on examples where $\Lambda=\Lambda_1 = 1$ and where $\gamma_a = \gamma_b = \gamma_c = 1$ outside of a region of compact support -- in other words cases where in the asymptotic region the only difference with the Atiyah-Hitchin metric occurs in the radial component of the metric $g_{r r}$ and which hence satisfy any physically reasonable boundary condition.   Roughly speaking, one may think of $\mu_0(r)$ as a ``mass function'' which tells one the amount of energy inside a given radius, although if one studied a broader class of examples than that below the other terms in (\ref{E2}) may be significant.

Turning now to the conditions we must impose in the interior of the space, note that  as $r \rightarrow r_0$ then $a(r)$ goes to zero and again we have a bubble of nothing solution.  When solving the constraint one finds a logarithmic divergence in $\mu_0(r)$ as $r \rightarrow r_0$ unless one insists that $\gamma^2_b(r_0) = \gamma^2_c(r_0)$.  To avoid a conical singularity as $r \rightarrow r_0$ one must take $W(r_0) = \gamma_a^{-2}(r_0)$ or equivalently
\be
\mu_0(r_0) = r_0 \gamma_b(r_0)\Big( 1 - \frac{\gamma_b^2(r_0)} {\gamma_a^{2}(r_0)}\Big)
\ee
A rather less obvious requirement for regularity requires that $b'(r_0) = c'(r_0)$, or given (\ref{Confdef}) that $\gamma_c'(r_0) =-\gamma_b'(r_0)$ as the necessary and sufficient condition, given the above conditions, to avoid the divergence of any component of the Riemann tensor in an orthonormal basis, and consequently the divergence of the square of the Riemann tensor and various other curvature invariants.   This condition intuitively is that statement that one may indeed reasonably approximate $b^2(r) = c^2(r) = C_0 r^2$ for some constant $C_0$ near the bubble (i.e. one really has an $S_2$, or with the identification II. an $RP_2$, whose volume grows as $r^2$).   
Given the above requirements we will have a regular bubble of nothing with size $\gamma_b(r_0) r_0$ (i.e. a metric of the form (\ref{nearmet}) with $r_0^2 d\Omega_2 \rightarrow \gamma^2_b(r_0) r^2_0 d \Omega_2$) and with $a'(r_0) = 2 \gamma_a(r_0)$.    Note that if one takes $\gamma_a^2(r_0) \ll \gamma_b^2(r_0)$ then necessarily $\mu_0(r_0)$ becomes negative and large in magnitude (presuming one does not take $\gamma_b(r_0)$ to be overly small) rather quickly as $\gamma_a(r_0)$ becomes relatively small. Recalling that the energy (\ref{E2}) is a function, aside from asymptotic contributions, of $\mu_0(\infty)$ and on rather general grounds one would expect positive contribution to $\mu(r)$ to be of order the size of the region where $(\gamma_a, \gamma_b, \gamma_c)$ are nontrivial, the above observations suggest one could find substantially negative energy configurations if $\gamma_a^2(r_0) \ll \gamma_b^2(r_0)$ and the $\gamma_i$ differ from unity only in some finite sized region.  

We have nearly assembled all the relevant pieces at this point.   It is reasonably clear, from examining several examples as well as the analytic form of $\mu_0(r)$ obtained by plugging (\ref{Confdef}) into (\ref{Const}) that any choice of $\gamma_i(r) $ (i.e. $\gamma_a(r), \gamma_b(r), \gamma_c(r)$) which are not approximately constant for $r \gg r_0$ only yield positive energy configurations; typically the integral that determines $\mu_0(r)$ will pick up positive contributions of order $r$ for any $r$ where the $\gamma_i$ are not constant.   It also seems difficult to find any negative energy solutions among choices of $\gamma_i$ simple enough to give $\mu_0(r)$ analytically.  Hence I will focus on numerical solutions where $\gamma_i$ are only nontrivial in some region around the bubble of nothing at $r = r_0$.  I then choose relative simple $C_1$ \footnote{Since the solution for $\mu_0(r)$ can be given in terms of at most first derivatives of the $\gamma_i$, one is free to make the solution as smooth as one likes at the cost of an arbitrarily small change in $\mu_0(r)$ and consequently the energy.} functions for $\gamma_i$ which are nonconstant only in a region of compact support ($r_0 < r < r_1$).  In terms of explicit numerical implementation, one may then numerically integrate out from $r_1$ to some $r_2$ where $r_2 \gg r_0$\footnote{As a matter of practice $r_2 = 10 r_0$ works tolerably well and $r_2 = 25 r_0$ very well.}.   For any greater $r$ one may approximate $a_0(r) = b_0(r) = r \sqrt{1-R_{11}/r}$ and $c_0(r) = -R_{11}/\sqrt{1-R_{11}/r}$ and obtain the explicit solution
\begin{eqnarray}
\mu_0(r) &=& \frac{(r - r_2) \sqrt{1-\frac{R_{11}}{r}} R^2_{11} (\Lambda^2 -\Lambda_1^2) }{(4 r -3 R_{11}) ( R_{11} - r_2) \Lambda} \nonumber \\
&+& \frac{\mu_0(r_2)  \sqrt{r_2} (r-R_{11})^{3/2} (4 r_2 -3 R_{11})}{\sqrt{r} (r_2 - R_{11})^{3/2} (4 r - 3 R_{11})}
\end{eqnarray}
and hence obtain $\mu_0(\infty)$ without needing to numerically integrate over extremely large ranges of $r$.   We are free to fix our gauge by choosing $\gamma_b = \Lambda$ and as discussed above I will focus on cases where $\Lambda = \Lambda_1 = 1$, although one seems to find qualitatively similar results in the case of $\Lambda_1 =1$ and $\Lambda$ of order one.   It is also useful to define a dimensionless measure of the energy
\be
\delta E = \frac{E - E_{AH}}{\vert E_{AH} \vert}
\ee
where, recall, the energy of the Atiyah-Hitchin solution is given by $E_{AH} =  -2 \pi^2 \kappa R^2_{11} V_T$. Choosing simple polynomials
\be \label{nransatz}
\gamma_a(r) = a_0 + a_1 r + a_2 r^2, \, \, \, \gamma_c(r) = c_0+c_1 r + c_2 r^2 + c_3 r^3
\ee
and solving for the constants $(a_0,a_1,a_2,c_0,c_1,c_2,c_3)$ via the boundary conditions discussed above  $\gamma_a(r_0) = \alpha, \gamma_c(r_0) = \Lambda, \gamma_c'(r_0) = 0, \gamma_a(r_1) = \Lambda, \gamma_c(r_1) = \Lambda_1, \gamma_a'(r_1) = \gamma_c'(r_1) = 0$ one obtains Figures 1-4.  It is easy to check with these boundary conditions the functions in (\ref{nransatz}) have no zeroes the range $r_0 < r < r_1$ and consequently the lifted orientifold may be smoothly distorted into these new solutions.  Before being plotted, each case was integrated at successive levels of numerical precision, $\mu_0(r)$ checked for regularity, and the positivity of $W(r)$ everywhere verified (to avoid complications with apparent horizons).  There does not appear to be anything particularly special about the form of $\gamma_a(r)$, $\gamma_c(r)$ chosen above.  One may, for example, add an additional power of $r$ to each expression in (\ref{nransatz}) to make the functions $C_2$ at $r = r_1$ and obtain qualitatively similar results.  Note all the plotted cases involve bubbles of the same size (i.e. $\Lambda/\Lambda_1$) as in the Atiyah-Hitchin solution.  One may consider cases of large or small bubbles, compared to the M-theory circle scale, but the boundary conditions above force significant gradients for $\gamma_c$ in that case and the resulting gradient energies always seem to result in $\delta E>0$.

\begin{figure}
\centering
    
\includegraphics[scale= 1.05]{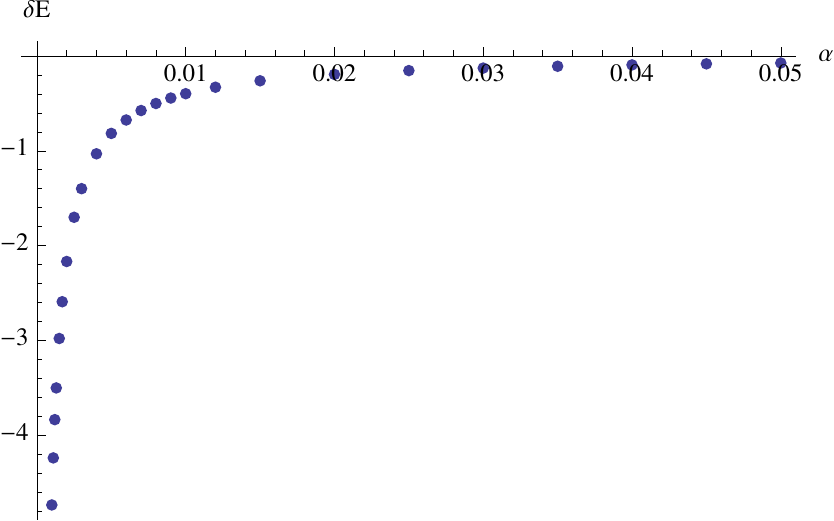}
	\caption{Masses for solutions versus $\alpha = \gamma_a(r_0)$ for $r_1/r_0 = 101/100$, $\Lambda = \Lambda_1 = 1$}
	\label{ex1fig}
	\end{figure}
	
	\begin{figure}
	\includegraphics[scale= 1.05]{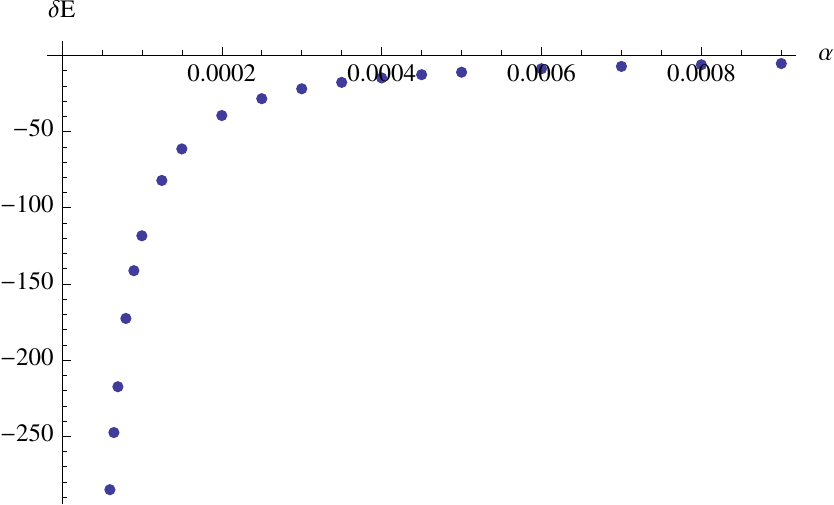}
	\caption{Masses for solutions versus $\alpha = \gamma_a(r_0)$ for $r_1/r_0 = 101/100$,  $\Lambda = \Lambda_1 = 1$and small $\alpha$}
	\label{ex2fig}
	\end{figure}
	
	\begin{figure}
	\includegraphics[scale= 1.05]{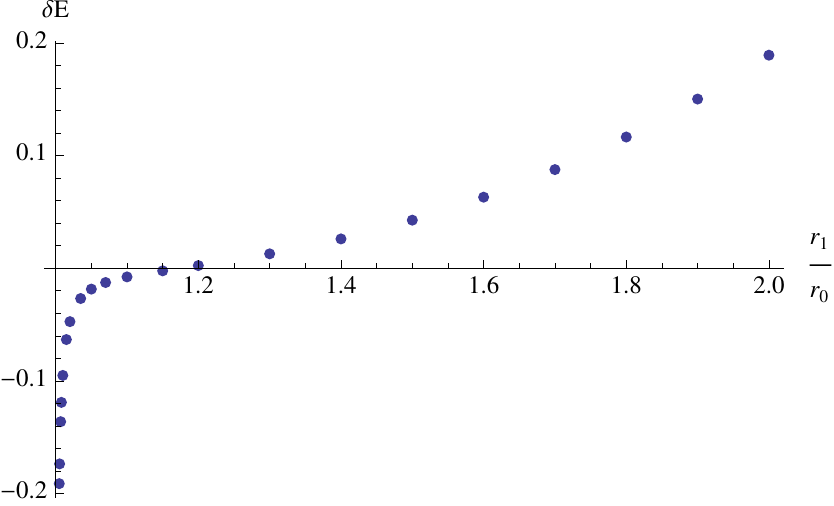}
	\caption{Masses for solutions versus $r_1/r_0$ for $\alpha = 1/25$ and $\Lambda = \Lambda_1 = 1$.}
	\label{ex3fig}
	\end{figure}
	
	\begin{figure}
	\includegraphics[scale= 1.05]{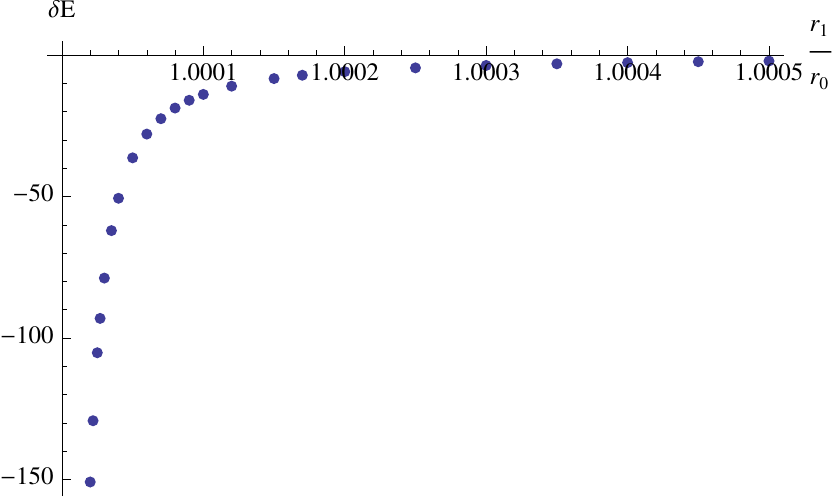}
	\caption{Masses for solutions versus $r_1/r_0$ for $\alpha = 1/25$ and $\Lambda = \Lambda_1 = 1$ as $r_1 \rightarrow r_0$.}
	\label{ex4fig}
	\end{figure}

The fact that the plotted examples include energies much more negative than the orientifold and the fact that the system has only a single scale, namely the size of the M-theory circle, which sets the scale of the orientifold mass strongly suggests there are solutions with arbitrarily negative energy.  Further the rapid plunge illustrated in Figures 1 and 2 as $\alpha$ becomes small for $r_1$ relatively close to $r_0$ or as $r_1$ approaches $r_0$ for relatively small $\alpha$ in Figures 3 and 4 appears to represent a clear numerical trend.   As suggested above, intuitively the behavior in Figures 1 and 2  this is exactly the kind of behavior one might expect given the fact that as $\gamma_a(r_0)/\gamma_b(r_0)$ becomes small $\mu_0(r_0)$ becomes rapidly more negative.   The behavior of decreasing energy as $r_1 \rightarrow r_0$ for fixed and relatively modest values of $\alpha$ is not as obvious and the best intuition I may offer the reader as of this writing is that small $\alpha$ results in a significantly negative $\mu_(r_0)$ and making this region small minimizes the amount of gradient energy resulting from the varying $\gamma$.  The fact that $\gamma_a(r)$ must interpolate between small values and $\Lambda$  (by $r= r_1$)  costs some amount of gradient energy.  Indeed, in some sense at leading order these effects cancel, for naively one might have estimated $\delta E \sim \alpha^{-2}$ whereas the actual results are far more modest.  As one might guess, this delicate balance suggests reproducing these results analytically is challenging and at least as of this writing an analytic approximation reproducing such numerical results remains elusive.  

On the other hand, one may obtain some analytic results for perturbative fluctuations around the lifted orientifold.  In particular one would like to know if the orientifold is energetically perturbatively stable, as one rather strongly suspects based on its supersymmetry.   Even this question of perturbative stability turns out to be surprising difficult as at least in the most obvious gauges (e.g. fixing $b(r) = b_0(r)$ as above) one finds a result for the energy quadratic in the action where one may not eliminate all terms linear in derivatives of the perturbation with respect to $r$ by integration by parts.  As a result, let us consider a more modest goal and limits our attention to cases $\Lambda = \Lambda_1 = 1$ and $\gamma_c = \Lambda_1 = 1$, as follows if, for example, as above one takes $\gamma_c$ to be a polynomial satisfying the above boundary conditions.   In this case if one takes $\gamma_a = 1+\epsilon \delta a(r)$, then $\delta E = \epsilon^2 \int_{r_0}^{\infty} [ f_0 \, \delta a^2(r) + f_1 (\delta a'(r))^2]+ \mathcal{O}(\epsilon^3)$ where $f_1$ is a positive definite function of $(a_0(r),b_0(r),c_0(r))$ and $f_0$ is likewise a function of $(a_0(r),b_0(r),c_0(r))$ and is positive definite except in a small range around the bubble ($r_0 \leq r \lesssim 1.2 r_0$) where it becomes modestly negative.   Following  the method introduced by Breitenlohner and Freedman \cite{BF} to show the perturbative stability of certain tachyons in anti-de Sitter space if one defines $\delta a = (r-r_c) \delta A(r)$ where $f_0(r_c) > 0$, say $r_c = 1.5 r_0$ for the sake of definiteness, upon an integration by parts one finds a positive definite expression for the energy in the region $r_0 < r < r_c$.

Given the above, the most likely dynamical evolution of any initial state of the specified form to evolve towards a similar state with lower energy, acquiring gravitational momentum along the way to conserve energy.   In terms of the dynamics, it is easy to check the simplest ansatz works, that is  adding a time direction to the spatial metric (\ref{met2}) and making each of the functions time dependent consistent solves the constraint and allows one to describe the evolution of any desired initial state.   Note that as an effectively two dimensional problem, it should be practical to analyze these dynamics numerically.  There then are three basic possibilities for the evolution--consisting of the bubble expanding, collapsing to a singularity or a black hole, or settling down to some static solution plus a shell of radiation.  As suggested above, bubbles that substantially expand or collapse appear to require the input of energy and hence appear dynamically unlikely scenarios.  The second proposition appears very dubious since negative masses repel each other and it tends to be quite difficult to form horizons with negative mass; the latter is a very good thing from a stringy point of view since high entropy objects with negative mass would be expected to be pair produced from the vacuum together with gravitational radiation at a Planckian rate.   It is not difficult to find static solutions which form a one-parameter generalization of the Atiyah-Hitchin metric parameterized by $a'(r_0)$ where $a(r)$ linearly approaches zero as $r \rightarrow r_0$.  All solutions besides the Atiyah-Hitchin metric, however, appear to either lead to singularities at finite $r$ or to an asymptotic metric which is not of the desired form even at leading order.  Based on the above, it appears dynamically one should expect the area of the bubble to stay nearly constant but for $a(r)$ to rapidly flatten near the bubble.  It turns out, as discussed below, this limit is a singular one, and since one should not expect the formation of a horizon, as discussed previously, one might expect to see a very strong violation of cosmic censorship in this context.

As one makes either $\alpha \rightarrow 0$ for fixed $(r_1/r_0,\Lambda, \Lambda_1)$ or $r_1/r_0 \rightarrow 1$ with $(\alpha,\Lambda, \Lambda_1)$ fixed one finds components of the Riemann tensor in an orthonormal basis (a convenient set is given by $e^{(1)}_{\alpha} = (W(r))^{-1/2} \delta_\alpha^r, e^{(2)}_{\alpha} = a(r) (\sigma_1)_{\alpha}$, and so forth) diverge and consequently so will a variety of curvature invariants, including the square of the Riemann tensor.   While the second case is not surprising, the first is rather less obvious and the only intuition I have to offer the reader at present is that one is approaching a solution where $a(r)$ vanishes faster than a linear function and hence rather than a regular $R_2 \times RP_2$ one might well expect a singularity.  Since $R_{11} \sim g_s l_s$ while the ten and eleven dimensional Planck scales are given by $l_p^{(10)} = g_s^{1/4} l_s$ and $l_p^{(11)} = g_s^{1/3} l_s$, respectively, at large $g_s$ there is a parameteric separation between $R_{11}$ and the Planck scale \cite{MarolfRossOrient},.   Consequently, at large string coupling, the $\alpha'$ corrections to even the solutions with energies much smaller than the orientifold are negligable.  Even if one cared to go beyond this regime, it is worth emphasizing that while $\alpha'$ corrections might be significant near the bubble, the mass is determined via the asymptotics where the curvature remains small in Planck units and classical general relativity remains reliable.   Hence it appears one is forced to conclude that while orientifolds appear to be perfectly acceptable in strict string perturbation theory, at significant string coupling they become badly unstable.   Since in realistic string compactifications $g_s$ tends to be of order one, essentially to avoid making the fine structure constant unnacceptably small, it seems clear one may not regard orientifolds as trustworthy objects in building string compactifications or a landscape.   On the other hand, in view of these results it is natural to ask if all string compactifications must be plagued by moduli or if this is merely a feature of those configurations understood to date. More broadly one might be concerned that via the various conjectured dualities between types of string theory, in particular the strong-weak string coupling duality between unoriented Type I theory (where it seems orientifolds are an inevitable result of T-duality \cite{DaiLeighPolchinski}) and heterotic theory, implies this pathology is a danger to all of string theory. Fortunately, provided one takes the interpretation that Type I theory and orientifolding are only well-defined in perturbation theory (where the instability of the orientifold becomes invisible), there seems to be little hard evidence for such a dire scenario.  In particular, it is by no means clear that the tests of this duality performed previously should have seen this instability.  However, a more complete exploration of these issues  would clearly be welcome.

%\smallskip
%%%%%%%%%%%%%%%%%%%%%%%%%%%%%%%%%%%%%%%%%%%%%%%%%%%%%%%%%%%%%%%%%%%%%%%%%%
\vskip .5cm
\centerline{\bf Acknowledgements}
\vskip .1cm
 It is a pleasure to thank D. Wesley, R. Myers, R. Mann, J. Polchinski,  F. Quevedo, and D. Marolf for useful comments and N. Turok for especially useful discussions.  This work was supported in part by the Natural Sciences \& Engineering Research
Council of Canada.   Research at Perimeter Institute is supported by the Government of Canada through Industry Canada and by the Province of Ontario through the Ministry of Research \& Innovation.

%%%%%%%%%%%%%%%%%%%%%%%%%%%%%%%%%%%%%%%%%%%%%%%


\begin{thebibliography}{99}

  \bibitem{MarolfRossOrient}
    D. Marolf and S. Ross, ``Stringy negative tension branes and the second law of thermodynamics,'' JHEP {\bf204} (2002)008 [arXiv: hep-th/0202091]

    
     \bibitem{Seiberg}
N.~Seiberg,
``IR dynamics on branes and space-time geometry,''
Phys.\ Lett.\ B {\bf 384}, 81 (1996),
hep-th/9606017.
%%CITATION = HEP-TH 9606017;%%

\bibitem{SW}
N.~Seiberg and E.~Witten,
``Gauge dynamics and compactification to three dimensions,''
hep-th/9607163.
%%CITATION = HEP-TH 9607163;%%

\bibitem{HP} A.~Hanany and B.~Pioline,
``(Anti-)instantons and the Atiyah-Hitchin manifold,''
JHEP {\bf 0007}, 001 (2000),
hep-th/0005160.
%%CITATION = HEP-TH 0005160;%%

\bibitem{Dorey}
N.~Dorey, V.~V.~Khoze, M.~P.~Mattis, D.~Tong and S.~Vandoren,
``Instantons, three-dimensional gauge theory, and the Atiyah-Hitchin
manifold,'' 
Nucl.\ Phys.\ B {\bf 502}, 59 (1997),
hep-th/9703228.
%%CITATION = HEP-TH 9703228;%%

\bibitem{GibbonsSUSYDomainWall} 
  G. W. Gibbons, H. Lu, C. N. Pope and K.S. Stelle,
 ``Supersymmetric domain walls from metrics of special holonomy,''
  Nucl.\ Phys.\ B {\bf 623}, 3 (2002)
  [hep-th/0108191].
  %%CITATION = HEP-TH/0108191;%%
  
  \bibitem{Wittenposenergy}
  E. Witten,
  ``A Simple Proof of the Positive Energy Theorem,''
  Commun.\ Math.\ Phys.\  {\bf 80}, 381 (1981).
  %%CITATION = CMPHA,80,381;%%
  
  \bibitem{ALFexamples} 
  G.W. Gibbons and C.N. Pope, ``Positive action theorems for
ALE and ALF spaces,'' ICTP/81/82-20, available from KEK;
C. LeBrun, ``Counterexamples to the generalised positive action
conjecture,'' Comm. Math. Phys. {\bf 118}, 591 (1988);
H. Nakajima, ``Self-duality of ALE Ricci-flat 4-manifolds and positive
mass theorem,'' Adv. Studies in Pure Math. {\bf 18-I}, 385 (1990);
M. Dahl, ``The positive mass theorem for ALE manifolds,'' Banach
Center Publ., {\bf 41}, Part 1;
A. Adams, J. Polchinski and E. Silverstein, ``Don't panic! Closed
string tachyons in ALE spacetimes,''  JHEP {\bf 0110}, 029 (2001)
  [hep-th/0108075].
  
  \bibitem{CdL}
   S.~R.~Coleman and F.~De Luccia,
``Gravitational Effects on and of Vacuum Decay,''
  Phys.\ Rev.\ D {\bf 21}, 3305 (1980).
  %%CITATION = PHRVA,D21,3305;%%

 \bibitem{KKbubbles}
D.~Brill and H.~Pfister, ``States of negative total energy in Kaluza-Klein theory,'' Phys. Lett. B {\bf 228}, 359 (1989); D.~Brill and G.~T.~Horowitz, ``Negative energy in string theory,'' ' Phys. Lett. B {\bf 262}, 437 (1991).


\bibitem{AdSSoliton}
 G.~T.~Horowitz and R.~C.~Myers,
``The AdS / CFT correspondence and a new positive energy conjecture for general relativity,''
  Phys.\ Rev.\ D {\bf 59}, 026005 (1998)
  [hep-th/9808079].
  %%CITATION = HEP-TH/9808079;%%
  
  \bibitem{CopseyHorowitzsol}
   K.~Copsey and G.~T.~Horowitz,
``Gravity dual of gauge theory on $S^2$ x $S^1$ x R,''
  JHEP {\bf 0606}, 021 (2006)
  [hep-th/0602003].
  %%CITATION = HEP-TH/0602003;%%

\bibitem{ReggeTeitelboim} 
  T.~Regge and C.~Teitelboim,
``Role of Surface Integrals in the Hamiltonian Formulation of General Relativity,''
  Annals Phys.\  {\bf 88}, 286 (1974).
  %%CITATION = APNYA,88,286;%%


\bibitem{SudarskyWald}
D.~Sudarsky and R.~M.~Wald,
``Extrema of mass, stationarity, and staticity, and solutions to the 
Einstein
Yang-Mills equations,''
Phys.\ Rev.\ D {\bf 46}, 1453 (1992);
%%CITATION = PHRVA,D46,1453;%%
R.~M.~Wald,
``The First law of black hole mechanics,''
arXiv:gr-qc/9305022.
%%CITATION = GR-QC 9305022;%%
\bibitem{CopseyMann} 
  K.~Copsey and R.~B.~Mann,
  ``States of Negative Energy and AdS(5) x S(5)/Z(k),''
  JHEP {\bf 0805}, 069 (2008)
  [arXiv:0803.3801 [hep-th]].
  %%CITATION = ARXIV:0803.3801;%%
  
 
\bibitem{WittenKK} 
  E.~Witten,
  ``Instability of the Kaluza-Klein Vacuum,''
  Nucl.\ Phys.\ B {\bf 195}, 481 (1982).
  %%CITATION = NUPHA,B195,481;%%
  
  \bibitem{BF}
  P.~Breitenlohner and D.~Z.~Freedman,
``Positive Energy in anti-De Sitter Backgrounds and Gauged Extended Supergravity,''
  Phys.\ Lett.\ B {\bf 115}, 197 (1982).
  %%CITATION = PHLTA,B115,197;%%
  
  \bibitem{DaiLeighPolchinski}
  J. Dai, R.G. Leigh, and J. Polchinski, ``New connections between string theories,''
Modern Physics Letters A 1989 04:21, 2073-2083 




\end{thebibliography}
\end{document}